\documentclass[prc,twocolumn,superscriptaddress,showpacs,floatfix,aps,10pt]{revtex4}

\pdfoutput=1

\usepackage{graphicx}
\usepackage{amsfonts}
\usepackage{dcolumn}
\usepackage{bm}
\usepackage{braket}
\usepackage{multirow} 
\usepackage{amsmath}
\usepackage{xcolor} 

\begin{document}

\date{\today}

\title{Gamow shell model description of $^4$He($^2$H,$^2$H) elastic scattering reactions}

\author{A. Mercenne}
\affiliation{Department of Physics and Astronomy, Louisiana State University, Baton Rouge, LA 70803, USA}

\author{N. Michel}
\affiliation{Grand Acc\'el\'erateur National d'Ions Lourds (GANIL), CEA/DSM - CNRS/IN2P3, BP 55027, F-14076 Caen Cedex, France}

\author{M. P{\l}oszajczak}
\affiliation{Grand Acc\'el\'erateur National d'Ions Lourds (GANIL), CEA/DSM - CNRS/IN2P3, BP 55027, F-14076 Caen Cedex, France}

\begin{abstract}
Structure of weakly bound/unbound light nuclei is often related to the low-energy decay channels involving composite particles like deuteron or $\alpha$-particle. These channels are essential to understand the appearance of light nuclei in the Big Bang nucleosynthesis or the helium fusion. We generalize the Gamow shell model (GSM) in coupled-channel (GSM-CC) representation to include reaction channels with the composite particles. In the core + valence particle formulation, this unified microscopic approach for structure and reactions involving weakly bound/unbound nuclei can be also applied to study low-energy properties of heavy nuclei. As the first application of this generalized GSM-CC approach, we describe the structure of $^6$Li and deuteron - $\alpha$-particle elastic scattering using the same effective Furutani-Horiuchi-Tamagaki (FHT) type nucleon-nucleon interaction. Asymptotically, the deuteron structure including its continuum is described using the N$^3$LO chiral force. The bulk of the data, including low-energy spectrum of $^6$Li, asymptotic normalization coefficients, and angular differential cross sections are satisfactorily described.
 \end{abstract}

\maketitle

\section{Introduction}
\label{sec1}

The comprehensive description of bound states, resonances and scattering many-body states within a single theoretical framework is one of the main goals of the nuclear theory. This is particularly important close to the edges of stability with respect to the particle emission where the coupling between resonant states and the non-resonant continuum is an important ingredient of the structure and the dynamics of the many-body system. Early attempts to reconcile nuclear structure with nuclear reaction theory \cite{fesh1,fesh2} lead to the development of the continuum shell model \cite{mahwei, barz} which in the recent applications \cite{benn00,rot05,volya05} evolved into the unified approach to nuclear structure and reactions. In this approach, one couples eigenstates of the phenomenological shell model hamiltonian with corresponding reaction channels to study the mutual influence of discrete and continuum many-body states on the level spectroscopy and the reaction cross-sections \cite{Oko03}. More recently, following the progress in the no-core shell model \cite{navr00} and the development of new methods based on the chiral effective field theory to devise the nucleon-nucleon and three-nucleon interactions \cite{int1,int2,PRC_N3LO,int3,int4}, the strategy pioneered by the continuum shell model has been extended to {\it ab initio} description of structure and reactions of light nuclei within the no-core shell model coupled with the resonating-group method (NCSM/RGM) \cite{navratil08,navratil11} and the no-core shell model with continuum (NCSMC) \cite{baroni13}. Other exact methods exist and have been applied successfully for light systems ($A\leq 5$). These include the Faddeev(-Yakubovsky) approach \cite{fad1,fad2,Elster_three_body}, the method of spherical harmonics \cite{HH}, or the Alt-Grassberger-Sandhas approach \cite{AGS1,AGS2}.

An alternative approach to unify the description of structure and reaction properties has been proposed with the open quantum system formulation of the shell model. Such a formulation is provided by the Gamow shell model (GSM) \cite{mic02,betan02,mic03,JPG_GSM_review} which offers the most general treatment of couplings between discrete and scattering states. The many-body states in GSM are given by the linear combination of Slater determinants defined in the Berggren ensemble \cite{berggren_1968} of single-particle states. In this way, the GSM is the tool for studies of bound and unbound many-body states and their decays. Most numerical applications of the GSM have been done by separating an inert core and  using the cluster orbital shell model (COSM) \cite{cosm} relative variables in valence space. In this way, the spurious center-of-mass excitations are removed. Moreover, an {\it ab initio} no-core formulation of the GSM has been recently developed to study resonant states in light nuclei \cite{PRC_George,PRL_Fossez}. 

For the description of scattering properties and reactions, it is convenient to formulate GSM in the representation of reaction channels. GSM in coupled-channel (GSM-CC) representation, which is based on the RGM, has been applied for various observables involving one-nucleon reaction channels, such as the excitation function and the proton/neutron elastic/inelastic differential cross-sections \cite{PRC_GSM_CC_Yannen,Mercenne16}, or low-energy proton/neutron radiative capture reactions \cite{PRC_GSM_CC_Kevin,Dong17}. Reaction channels in these processes are given by the initial/final GSM eigenvectors of $(A-1)$-body system coupled to proton/neutron in continuum states. All resulting $A$-body wave functions are fully antisymmetrized and the separation of core and valence particles allows to apply the GSM-CC approach for small number of valence particles in medium-heavy and heavy nuclei. 
 
The main purpose of the present paper is to extend the GSM-CC approach to reactions with cluster reaction channels, such as $^2$H or $^{4}$He reaction channels. Such channels appear often at low excitation energies in $p$- and $(sd)$-shell nuclei and therefore, have great importance for decay properties and low-energy transfer or radiative capture reactions in these nuclei. In medium mass nuclei, the low-energy reactions with deuteron or $\alpha$-particle lead to the formation of nuclear aggregate at higher excitation energies where the density of states is significant. Here the practical restriction in GSM-CC to valence space effective interactions may be a convenient way to describe these low-energy reactions microscopically. 

As  the first application and a testing ground of GSM-CC with cluster reaction channels, we shall discuss the structure of $^6$Li and the elastic scattering of deuteron on $\alpha$-particle at low center-of-mass energies. This problem was analyzed before in RGM formalism \cite{rgm1}, and more recently in the NCSM/RGM \cite{navratil11a} and NCSMC \cite{PRL_Hupin} formalisms using nucleon-nucleon and three-nucleon chiral effective interactions. The comparison between an {\it ab initio} NCSMC and a more phenomenological GSM-CC for the same system provides the useful insight into the reliability of the latter approach which in principle could be applied also in heavier systems. 

$^6$Li in GSM-CC approach is described as a $^4$He core and two valence particles interacting with the finite-range FHT type interaction \cite{1979Furutani,1980Furutani,PRC_FHT_Yannen}. The coupled-channel equations of the GSM-CC are solved using Berggren basis expansion method. Deuteron which is a weakly bound nucleus, requires careful treatment of continuum to include its polarization and virtual breakup, i.e the different eigenvalues of the intrinsic deuteron Hamiltonian in the collision with $\alpha$-particle. The deuteron structure is described in Berggren basis using the N$^3$LO chiral interaction \cite{PRC_N3LO}. 

The GSM-CC formalism for the antisymmetrized cluster channel states is presented in Chapter \ref{sec2}. 
The model space and the nucleon-nucleon interaction used in this work are detailed in Chapter \ref{sec3}. Results of the GSM-CC calculation for the elastic phase-shifts, excitation function and differential cross sections are discussed in Chapter \ref{sec4}. Finally, main conclusions of this work are summarized in Chapter \ref{sec5}.

\section{Description of nuclear reactions in GSM-CC}
\label{sec2}

We will describe in this section the theory of GSM-CC with cluster projectiles. 
GSM-CC has already been defined for the case of one-nucleon projectile in Ref.\cite{PRC_GSM_CC_Yannen}, so that we will mainly concentrate on the differences between one and many-body clusters.
For this, the numerical method to generate cluster wave functions with well defined intrinsic and center of mass parts will be presented in Sec. \ref{sec2.1} . 
The basis-generating potential of the center of mass part of projectiles, based on the cluster approximation of the used Hamiltonian, will be explicited in Sec. \ref{sec2.2}.
The coupled-channel equations of the GSM-CC model will then be formally derived in Sec. \ref{sec2.3} , where antisymmetry requirements between target and projectile will be emphasized.
A numerical method to solve GSM-CC coupled-channel equations with the Berggren will be described afterwards in Sec. \ref{numerical_resolution} , 
with which, in particular, direct integration in coordinate space is replaced by matrix diagonalization and linear systems to solve.

\subsection{Cluster states definition in relative and laboratory frames}
\label{sec2.1}

Projectile states read:
\begin{equation}
  \ket{\Psi_{ p }^{ J_{p}}} = { \left[ \ket{K_{ CM } , L_{CM}} \otimes \ket{K_{\text{int}} , J_{\text{int}}} \right] }_{ M_{p}}^{ J_{p}} \;
  \label{projectile_state}
\end{equation}
where $K_{CM}$ and $L_{CM}$ are respectively the linear momentum and angular momenta of the center of mass, $K_{\text{int}}$ is the intrinsic linear momentum,
and $J_{\text{int}}$ represent the intrinsic angular momenta so that we have $\mathbf{ J_{ p }} = \mathbf{ J_{ int }} + \mathbf{ L_{CM}}$.

Composite states are then built from the antisymmetrized tensor product of target and projectile states. 
For antisymmetry to be fulfilled, one expands both target and projectile in the same complete basis of Slater determinants.
As the target state was generated by a GSM calculation, it is expanded with Slater determinants by construction. 
Equation(\ref{projectile_state}) must be then expanded in the basis of Slater determinants used for the target:
\begin{equation}
  \ket{\Psi_{p}^{ J_{p}}} = \sum_{ n } C_{n} \ket{SD_{ n }} \; ,
  \label{Berggren_naive_expansion}
\end{equation}
where the Slater determinants are constructed from the single-particle states of the Berggren ensemble.
However, the overlap $\braket{\Psi_{p}^{ J_{p}} | { \Psi }_{ p' }^{ { J }_{ p' }} } = \delta(K_{CM} - {K'}_{CM})$ 
is difficult to treat numerically, because the treatment of the Dirac delta function would require an extremely fine discretization of the continuum for the center of mass/intrinsic separation of 
$\ket{\Psi_{p}^{ J_{p}}}$ in Eq. (\ref{projectile_state}) at large distances. 

Consequently, one has to proceed indirectly.
As reactions are localized close to the target, the wave function of the projectile can be approximated by a bound state wave function, so that we can use the harmonic oscillator (HO) basis instead.
Let us define the HO projectile state as:
\begin{equation}
  \ket{\Psi_{p}^{ J_{p}}}^{ \text{HO} } = { \left[ \ket{N_{ CM } , L_{CM}}^{ \text{HO} } \otimes \ket{K_{\text{int}} , J_{\text{int}}}^{ \text{HO} } \right] }_{ M_{p}}^{ J_{p}} \; .
  \label{HO_rel_cm}
\end{equation}
where $\ket{N_{ CM } , L_{CM}}^{ \text{HO} }$ is a center-of-mass harmonic oscillator state and $\ket{K_{\text{int}} , J_{\text{int}}}^{ \text{HO} }$ is an intrinsic deuteron state expanded on a basis of harmonic oscillator states.

In order to calculate the intrinsic part of $\ket{\Psi_{p}^{ J_{p}}}^{ \text{HO} }$ in Eq. (\ref{HO_rel_cm}), 
the intrinsic Hamiltonian $\hat{H}_{\text{int}}$ is firstly diagonalized on a Berggren basis of only bound and scattering states since the deuteron does not have resonances. This guarantees a good asymptotic behavior of relative scattering deuteron states.

The absence of resonance states in the single-particle basis poses no problem therein because the used Berggren basis is complete for the calculation of bound and scattering intrinsic deuteron states. Indeed, deuteron possesses no low-lying resonance states, so that the Berggren basis contour can efficiently be discretized with the Gauss-Legendre quadrature.

This is a one-body problem in our case as we just have deuteron projectiles, so that $\hat{H}_{\text{int}}$ is a small matrix therein and can be exactly diagonalized.  
This provides with the relative deuteron eigenstates $\ket{K_{\text{int}} , J_{\text{int}}}$ generated by $\hat{H}_{\text{int}}$ in the relative Berggren basis.
They are projected on a basis of harmonic oscillator states in order to provide with the $\ket{K_{\text{int}} , J_{\text{int}}}^{\text{HO}}$ states.

The $\ket{\Psi_{p}^{ J_{p}}}^{ \text{HO} }$ states of Eq. (\ref{HO_rel_cm}) have to be expressed in laboratory coordinates to be able to use them in a shell model formalism. 
For this, one can apply the Brody-Moshinsky transformation \cite{brody_moshinsky_coeffs} to the relative + center-of-mass two-body wave functions formed by the deuteron states of Eq. (\ref{HO_rel_cm}) as they are two-body systems. 

Consequently, the coefficients of the Slater determinant expansion of $\ket{\Psi_{p}^{ J_{p}}}^{ \text{HO} }$ in Eq. (\ref{HO_rel_cm}) are straightforward to obtain:
\begin{equation}
  \ket{\Psi_{p}^{ J_{p}}}^{ \text{HO} } = \sum_N C_N^{ \text{HO} } \ket{SD_{N}}^{ \text{HO} }
  \label{HO_sd}
\end{equation}

We can now express the projectile state in the Berggren basis:
\begin{equation}
  \ket{\Psi_{p}^{ J_{p}}}^{ \text{HO} } = \sum_{ n } C_{n} \ket{SD_{ n }} \; .
  \label{HO_Berggren_sd}
\end{equation}
where the $C_{ n }$ coefficients come from a direct expansion of the HO basis to the Berggren basis:
\begin{eqnarray}
 C_{n} &=& { \braket{ { SD }_{ n } | \Psi_{p}^{ J_{p}} } }^{ \text{HO} } \nonumber \\ 
 &=& \sum_N C_N^{ \text{HO} } { \braket{ SD_{ n } |SD_N} }^{ \text{HO} } \; .
\end{eqnarray}
Note that the $\braket{SD_{ n } | SD_N}^{ \text{HO} }$ overlap is easily computed even though $\ket{SD_{ n }}$ and $\ket{SD_{ N }}^{ \text{HO} }$ are built from different one-body basis states, 
because the two latter Slater determinants bear only one proton and one neutron state, so that $\braket{ SD | SD}^{ \text{HO} }$ is a product of proton and neutron overlaps:
\begin{equation}
  \braket{ SD | SD}^{ \text{HO} } = \braket{ \phi^{(p)} | \phi^{(p)}}^{ \text{HO} } \braket{ \phi^{(n)} | \phi^{(n)}}^{ \text{HO} }
  \label{SD_overlap}
\end{equation}
where $\ket{\phi^{(p(n))}}$ and $\ket{{\phi^{(p(n))}}}^{ \text{HO} }$ denote the proton (neutron) one-body states of the $\ket{SD}$ and $\ket{SD}^{ \text{HO} }$ Slater determinants, respectively.
Aside from the HO projection of the intrinsic state $\ket{K_{\text{int}} , J_{\text{int}}}$, Eqs.(\ref{Berggren_naive_expansion},\ref{HO_Berggren_sd}) differ also by their center of mass part,
as it is a Berggren center-of-mass state in Eq. (\ref{projectile_state}) and a HO state in Eq. (\ref{HO_rel_cm}). 
As the center-of-mass asymptote of composite states will be provided by the integration of coupled-channel equations, defined in coordinate space, having a localized center-of-mass state in 
Eq. (\ref{HO_Berggren_sd}) creates no problem. Those projectile states will later be used as a basis in which the reaction potentials, that need to be integrated, will be calculated. The use of an HO basis expansion for the c.m. part is thus justified since it provides a localized basis.

This procedure can be easily generalized to cluster bearing more than two protons or neutrons 
by considering the explicit expansion of Slater determinants in terms of non-antisymmetrized tensor products of proton or neutron one-body states.
As clusters will not have more than two protons or neutrons in practice, and $\alpha$ clusters being the heaviest projectiles that we plan to consider, 
a direct calculation of the overlap of Slater determinants will not be a caveat for future calculations involving clusters other than deuterons.
One might argue that the relative + center-of-mass treatment used in this section cannot be generalized to heavier clusters, as the Brody-Moshinsky transformation is valid only for two-nucleon systems.
This problem can be solved for projectiles of 3 and 4 nucleons.
On the one hand, if break-up can be neglected, which is the case at low energy due to the well bound character of considered projectiles,
all calculations of cluster states can be effected with the HO basis in $N \hbar \omega$ spaces, where the separation of relative and center-of-mass degrees of freedom can be done exactly.
On the other hand, the inclusion of break-up would demand the use of Jacobi coordinates to calculate the relative scattering states of considered clusters, which is feasible with three-nucleon systems in particular.

\subsection{Berggren basis of center-of-mass cluster states}
\label{sec2.2}
  {
    The $\ket{\Psi_{p}^{ J_{p}}}^{ \text{HO} }$ state of Eqs.(\ref{HO_rel_cm},\ref{HO_sd}) will be used to build the coupled-channel Hamiltonian. 
    However, in order to properly deal with the asymptotic behavior of the scattering states solutions of the coupled-channel equations of the Hamiltonian,
    we have to compute projectile states of the form given by Eq. (\ref{projectile_state}), where the center-of-mass part is generated by a finite-range potential.
    These states will be used theoretically to formally derive the Hamiltonian coupled-channel equations, on the one hand, and will be used numerically to expand the solutions of the coupled-channel equations, on the other hand.

    We will derive a Hamiltonian generating the latter states from the composite A-body Hamiltonian written in laboratory coordinates.
    The composite COSM Hamiltonian cannot be used for this two-nucleon projectile, as it is defined for a wave function where core particles are present. 
    The Hamiltonian in laboratory coordinates reads:
    \begin{equation}
      \hat{ H } = \sum_{ i } \frac{ { \mathbf{p_{ i,lab }} }^{ 2 }}{ 2 { m }_{ i }} + \sum_{ i < j } { \hat{ V }}_{ ij }
    \end{equation}
    where $i , j$ cover all nucleons and $\hat{ V }_{ ij }$ is the nucleon-nucleon interaction in laboratory coordinates.
    Let $\hat{U}_{ i }^{T}$ be the mean-field created by all target nucleons:
    \begin{equation}
      \sum_{ j \in \text{T}} { \hat{ V }}_{ ij } \rightarrow \hat{U}_{ i }^{T}.
    \end{equation}
    Thus, neglecting inter-nucleon interactions between target and projectile, the projectile Hamiltonian $\hat{ H }_{ p }$ in the laboratory frame can be defined:
    \begin{equation}
      \hat{ H }_{ p } = \sum_{ i \in p } \left(  \frac{\mathbf{p_{ i,lab }}^{ 2 }}{ 2m } + \hat{U}_{ i }^{T} \right) + \sum_{ i < j \in p } \hat{ V }_{ ij } \label{Hp} \; .
    \end{equation}
    Linear momenta of valence particles are identical in laboratory and COSM coordinates. 
    Moreover, one can show that the core corrections issued from potentials are second-order when one replaces laboratory radial coordinates by COSM radial coordinates, so that they can be neglected \cite{COSM_ref}.
    Indeed, the described procedure therein just aims at producing the optimal basis potential to generate $\ket{K_{CM} , L_{CM} , K_{\text{int}} , J_{\text{int}}}$ projectile states and does not affect the many-body Hamiltonian.
    Therefore $\hat{ H }_{ p }$ in COSM coordinates reads the same as in Eq. (\ref{Hp}). 
    Hence, from now on, we will use COSM coordinates ${ \left( \bold{ { p }_{ i } }, \bold{ { r }_{ i } } \right) }$ only.
    Then, we can write :
    \begin{eqnarray}
      \hat{ H }_{ p } &=& \sum_{ i \in p } \left( \frac{ \mathbf{p_{ i }}^{ 2 }}{ 2m } + \hat{U}_{ i }^{T} \right) + \sum_{ i < j \in p } \hat{ V }_{ ij } \nonumber \\
                     &=& \sum_{ i \in p } \frac{ { \left( \mathbf{p_{ i }} - \frac{ 1 }{ a } \mathbf{P_{CM}} \right) }^{ 2 }}{ 2m } + \sum_{ i < j \in p} \hat{ V }_{ ij } \nonumber \\ 
                     &+& \frac{ \mathbf{P_{CM}}^{ 2 }}{ 2 M_{p}} + \sum_{ i \in p } \hat{U}_{ i }^{T} \label{Hp_definition}
    \end{eqnarray}
    where
    \begin{equation}
      \mathbf{P_{CM}} = \sum_{ i \in p } \mathbf{p_{ i }} \label{PCM}
    \end{equation}
    $a $ the number of nucleon in the projectile, $m$ the nucleon mass, and $M_{p}$ the mass of the projectile.
  Assuming cluster approximation, i.e implying $\mathbf{r_i} \simeq \mathbf{R_{CM}}$, the central part of the mean-field $\hat{U}_{ CM }^{(c)} ( \mathbf{ { R } _{ CM }})$ created by all target nucleons can be approximated by:
    \begin{eqnarray}
      \!\!\!\!\!\!\!\!\hat{U}_{ CM }^{(c)} ( \mathbf{R_{CM}}) &\simeq& \sum_{ i \in p } \hat{U}^{T(c)}_{i}  (\mathbf{R_{CM}}) \nonumber \\
    &\simeq& a_p~\hat{U}^{T(c)}_{p} ( \mathbf{R_{CM}}) + a_n~\hat{U}^{T(c)}_{n} ( \mathbf{R_{CM}}) \label{UCM_central}
    \end{eqnarray}
    where $a_p$ and $a_n$ are the number of protons and neutron of the projectile respectively.
    The spin-orbit part of the mean-field $\hat{U}_{ CM }^{ (so) } ( \mathbf{R_{CM}})$ is calculated through a similar averaging procedure, 
    as the cluster approximation also implies that $\mathbf{l_i} \simeq \mathbf{L_{CM}} / a$ and $\displaystyle \sum_{ i \in p } \mathbf{s_i} \simeq \mathbf{J_{\text{int}}}$:
    \begin{eqnarray}
    && \sum_{ i \in p } \hat{U}^{T(so)}_{i}  ( \mathbf{R_{CM}}) ~ \mathbf{l_i} \cdot \mathbf{s_i} \nonumber \\
    &=& \hat{U}_{ CM }^{ (so) } ( \mathbf{R_{CM}}) ~ \sum_{ i \in p } \mathbf{l_i} \cdot \mathbf{s_i} \nonumber \\
    &=& \frac{1}{a} ~ \hat{U}_{ CM }^{ (so) } ( \mathbf{R_{CM}}) ~ \mathbf{L_{CM}} \cdot \mathbf{J_{int}} \label{UCM_so}
    \end{eqnarray}
    where $\hat{U}_{ CM }^{ (so) }$ is the average of all $\hat{U}^{T(so)}_{i}$ potentials:
    \begin{eqnarray}
     \hat{U}_{ CM }^{ (so) } ( \mathbf{R_{CM}}) &=& \frac{a_p}{a}~\hat{U}^{T(so)}_{p} ( \mathbf{R_{CM}}) \nonumber \\
                                           &+& \frac{a_n}{a}~\hat{U}^{T(so)}_{n} ( \mathbf{R_{CM}}) \label{UCM_so_average}
    \end{eqnarray}
    The $\hat{U}_{ CM } ( \mathbf{R_{CM}})$ potential, generating the $\ket{K_{CM} , L_{CM}}$ center-of-mass states thus reads from Eqs.(\ref{UCM_central},\ref{UCM_so}):
    \begin{eqnarray}
     \hat{U}_{ CM } ( \mathbf{R_{CM}}) &=& \hat{U}_{ CM }^{(c)} ( \mathbf{R_{CM}}) \nonumber \\
                                      &+& \frac{1}{a} ~ { \hat{U}_{ CM }^{ (so) }} ( \mathbf{R_{CM}}) ~ \mathbf{L_{CM}} \cdot \mathbf{J_{int}} \label{UCM}.
    \end{eqnarray}

    Consequently, $\hat{ H }_{ p }$ reads:
    \begin{equation}
      \hat{ H }_{ p } = \hat{ H }_{ int } + \hat{ H }_{ CM }
      \label{projectile_hamiltonian}
    \end{equation}
    with:
    \begin{equation}
      \hat{ H }_{ int } = \sum_{ i \in p } \frac{ { \left( \mathbf{p_{ i }} - \frac{ 1 }{ a } \mathbf{P_{CM}} \right) }^{ 2 }}{ 2m } + \sum_{ i < j \in p } { \hat{ V }}_{ ij } \label{H_int}
    \end{equation} 
    and
    \begin{equation}
      \hat{ H }_{ CM } = \frac{ { \mathbf{P_{CM}} }^{ 2 }}{ 2 M_{p}} + \hat{U}_{ CM } \label{HCM}.
    \end{equation}

     The mean-field approximation described in Eqs.(\ref{UCM_central},\ref{UCM_so}) insures that $\hat{U}_{ CM } ( \mathbf{R_{CM}})$ recaptures the features of the Hamiltonian of Eq. (\ref{Hp})
     for the $\ket{K_{CM} , L_{CM}}$ center-of-mass states at cluster approximation level. 
     It is thus possible to calculate a Berggren basis of center-of-mass $\ket{K_{CM} , L_{CM}}$ states, as these states are formally identical to one-body states.
     Moreover, even though cluster approximation is no longer valid at cluster break-up, the latter potential can still be used therein as it provides with a complete set of $\ket{K_{CM} , L_{CM}}$ center-of-mass states.
  }

\subsection{Formulation of Hamiltonian coupled-channel equations}
\label{sec2.3}
  {
    We then develop coupled-channel equations for multinucleon projectiles.
    We consider an $A$-body state decomposed in reaction channels:
    \begin{equation}
      \ket{\Psi_{ M_{ A }}^{J_{ A }} } = \sum_{ c } \int_{ 0 }^{ +\infty } \ket{{ \left( c , R \right) }_{ M_{ A }}^{J_{ A }} } \frac{ { u }_{ c } (R) }{ R } { R }^{ 2 } ~ d R \; ,
      \label{scat_A_body_compound}
    \end{equation}
    where the center-of-mass subscript is now dropped for CM radial coordinates and angular momenta for convenience, 
    $u_{ c } (R)$ is the radial amplitude of the $c$ channel to be determined, 
    $J_A$ and $M_A$ are the total angular momentum and total angular momentum projection of the $A$-body state
    and
    \begin{align}
      \ket{ \left( c , R \right)} & = \hat{ \mathcal{A}} \ket{ \{ \ket{\Psi_{ T }^{J_{ T }} } \otimes \ket{R ~ L ~ J_{\text{int}} ~ J_{p}} \}_{ M_{ A }}^{J_{ A }}} \nonumber \\ 
      & = \hat{ \mathcal{A} } \ket{ { \Phi }_{ A } },
      \label{channel}
    \end{align} 
    where the channel index $c$ stands for the $\{ A - a , { J }_{ T } ; a , { L } , J_{\text{int}}, J_{p}\}$ quantum numbers, 
    is an antimmetrized tensor product of the $\ket{\Psi_{ T }^{J_{ T }} }$ target state, and projectile channel state $\ket{R ~ L ~ J_{\text{int}} ~ J_{p}}$,
    where angular momentum couplings read $\mathbf{ J_{ p }} = \mathbf{ J_{ int }} + \mathbf{L}$ and  ${ \mathbf{ J_A} = \mathbf{J_p} + \mathbf{ J_{T} } }$.
    Due to the use of COSM, the relative motion between the two clusters is already defined within the projectile state, and it is not necessary to introduce a delta function in Eq.(\ref{channel}), as it can be commonly seen in RGM and NCSM/RGM \cite{navratil08} where laboratory coordinates are considered.

    The coupled-channel equations can then be formally derived from the Schr{\"o}dinger equation: $H \ket{\Psi_{M_{ A }}^{J_{ A }}} = E \ket{\Psi_{M_{ A }}^{J_{ A }}}$, as:
    \begin{equation}
      \!\sum_{c}\int_{0}^{\infty}  \!\!\! R^{ 2 } \left( H_{ cc' } (R , R') - E N_{ cc' } (R , R') \right) \frac{ { u }_{ c } (R) }{ R } = 0	
      \label{cc_cluster_eq}
    \end{equation}
    where
    \begin{align}
      & H_{ cc' } (R,R') = \bra{ (c,R) } \hat{ H } \ket{(c',R') } \label{h_cc_compound} \\
      & N_{ cc' } (R,R') = \braket{ (c,R) | (c',R') } \label{n_cc_compound}
    \end{align}
    
    Due to the decoupling of the target and projectile at high energy, it is more convenient to rewrite the Hamiltonian $\hat{ H }$ by introducing the target Hamiltonian $\hat{ H }_{ T }$:
    \begin{equation}
      \hat{ H }_{ T } = { \hat{ T }}_{ T } + \hat{U}_{ basis }^{ T }  + { ({ \hat{ V }}_{ res } - \hat{U}_{ 0 }) }^{ A-a }
      \label{hamiltonian}
    \end{equation}
    where $(\hat{ V }_{ res } - \hat{U}_{ 0 })^{ A-a }$ is the part of $\hat{ V }_{ res } - \hat{U}_{ 0 }$ acting on the $(A-a)$-body target state, 
    and where $\hat{ T }_{ T }$, $\hat{U}_{ basis }^{ T }$ are the target kinetic part of the Hamiltonian $\hat{ H }_T$ and the associated one-body basis potential, respectively. 
    The action of target and projectile Hamiltonians $\hat{ H }_{ T }$ and $\hat{ H }_{ p }$ (see Eq. (\ref{projectile_hamiltonian})) on $A$-body states is effected by considering non fully antisymmetrized $A$-body states:
    \begin{align}
      \hat{ H }_{ T } \left( \ket{{ \Psi }_{ T }} \otimes \ket{\Psi_{p}} \right) & = (\hat{ H }_{ T } \ket{{ \Psi }_{ T }} \otimes \ket{\Psi_{p}}) \label{HT_action}  \\
      \hat{ H }_{ p } \left( \ket{{ \Psi }_{ T }} \otimes \ket{\Psi_{p}} \right) & = \left( \ket{{ \Psi }_{ T }} \otimes \hat{ H }_{ p } \ket{\Psi_{p}} \right) \label{Hp_action}
    \end{align}
    Note that the target and projectile states above are already antisymmetrized. Thus, we can write the Hamiltonian as:
    \begin{equation}
      \hat{ H } = \hat{ H }_{ T } + \hat{ H }_{ p } + \hat{ H }_{ Tp }		
      \label{new_hamiltonian}
    \end{equation}
    where $\hat{ H }_{ Tp } = \hat{ H } - \hat{ H }_{ T } - \hat{ H }_{ p }$ by definition.
    
    The matrix elements $H_{ cc' }^{ J_{ A } M_{ A }} (R , R')$ can be formally expanded in a basis of ${ \ket{ (C,N) } = \ket{ N L { J }_{ int } { J }_{ p } } }$ states, 
    where $\hat{ H }_{CM} \ket{N ~ L } = E_{CM} \ket{N ~ L}$. 
    Note that here $N$ refers to a cluster center-of-mass Berggren basis state, arising from the discretization of the cluster center-of-mass complex contour, and not to a cluster center-of-mass HO state.
    The eigenbasis of $\hat{ H }_{CM}$ is indeed more convenient to formally derive the coupled-channel equations associated to clusters.
    Using now the following expansion : 
    \begin{equation}
      \ket{ (C,R) } = \sum_{N} \frac{ { U }_{ N }(R) }{ R } \ket{ (C,N) } 
      \label{}
    \end{equation}
    we can write :
    \begin{eqnarray}
      &&H_{ cc' } (R , R') = \nonumber \\
      && \sum_{ N , N'} H_{ cc' }^{ J_{ A } M_{ A }} (N , N') \frac{ U_{ N} (R) }{ R } \frac{ U_{ N'} (R') }{ R' } 
      \label{matrix_elmts_h_k}
    \end{eqnarray}
    which can be decomposed in four sums:
    \begin{eqnarray}
      &&H_{ cc' } (R , R') = \nonumber \\
      &&\sum_{ \stackrel{ N  \leq N_{ max }} { N' \leq N_{ max }} } H_{ cc' }^{ J_{ A } M_{ A }} (N , N') \frac{ { U_{ N} } (R) }{ R } \frac{ U_{ N'} (R') }{ R' }  \nonumber \\
      & + &  \sum_{ \stackrel{ N  \leq N_{ max }} { N' > N_{ max }} } H_{ cc' }^{ J_{ A } M_{ A }} (N , N') \frac{ { U_{ N} } (R) }{ R } \frac{ U_{ N'} (R') }{ R' }  \nonumber \\
      & + &  \sum_{ \stackrel{ N > N_{ max }} { N' \leq N_{ max }} } H_{ cc' }^{ J_{ A } M_{ A }} (N , N') \frac{ { U_{ N} } (R) }{ R } \frac{ U_{ N'} (R') }{ R' }  \nonumber \\ 
      & + &  \sum_{ \stackrel{ N > N_{ max }} { N' > N_{ max }} } H_{ cc' }^{ J_{ A } M_{ A }} (N , N') \frac{ { U_{ N} } (R) }{ R } \frac{ U_{ N'} (R') }{ R' }  
      \label{matrix_elmts_h_k_decomposed}
    \end{eqnarray}
    where $N_{max}$ is considered sufficiently large so that the antisymmetrization and thus $\hat{H}_{Tp}$ can be neglected for $N > N_{max}$ or $N' > N_{max}$.
    The expansion wave functions ${ { U }_{ N }(R) }$ depend on the relative angular momentum ${ L }$ and ${ { J }_{ \text{int} } }$, but since those numbers are already included in the channel index ${ c }$, this notation has been dropped for convenience.
    This property arises from the fact that the effective nuclear interaction used in the GSM target is defined in a finite model space, so that its high energy matrix elements vanish. 
    The use of the optimal center-of-mass potential of Eq. (\ref{UCM}) to generate $\ket{N ~ L}$ states implies that $N_{ max }$ does not have to be extremely large.

    The first term in Eq. (\ref{matrix_elmts_h_k_decomposed}) is a finite sum and can be calculated numerically using standard shell model formulas.
    The second sum with $N \leq N_{ max }$ and $N' > N_{ max }$ will be shown to be equal to zero:     
    \begin{eqnarray}
      &&H_{ cc' }^{ J_{ A } M_{ A }} \left( N , N' \right)  \nonumber \\ 
      &=&\bra{\Phi_{A}} \hat{ \mathcal{A}} \hat{ H } \hat{ \mathcal{A}} \ket{\Phi'_{A}} \nonumber \\
      &=& \bra{\Phi_{A}} \hat{ H }_{ T } + \hat{ H }_{ p } + \hat{ \mathcal{A}} \hat{ H }_{ Tp } \hat{ \mathcal{A}} \ket{\Phi'_{A}} \nonumber \\
      &=& \bra{\Phi_{A}} E_{ CM } + E_{ T } + E_{ int } \ket{\Phi'_{A}} + \bra{\Phi_{A}}\hat{ \mathcal{A}} \hat{ H }_{ Tp } \hat{ \mathcal{A}} \ket{\Phi'_{A}} \nonumber \\ 
      &=&  (E_{ CM } + E_{ T } + E_{ int }) ~ \delta_{ cc' } \delta_{ N N'} + \bra{\Phi_{A}} \hat{ \mathcal{A}} \hat{ H }_{ Tp } \hat{ \mathcal{A}} \ket{\Phi'_{A}} \nonumber \\
      &=& 0 \nonumber \\
      \label{expansion_second_sum}
    \end{eqnarray}
    where 
    \begin{eqnarray}
      \ket{\Phi_{A}}  &=& \left[ \ket{{ \Psi_T  }^{ J_T }}  \otimes \ket{N  ~ { L  } ~ { J  }_{ int } ~ { J  }_{ p } ~ { M  }_{ p }} \right]^{J_{ A }}_{M_{ A }} \label{Phi} \\
      \ket{\Phi'_{A}} &=& \left[ \ket{{ \Psi'_T }^{ J'_T }} \otimes \ket{N' ~ { L' } ~ { J' }_{ int } ~ { J' }_{ p } ~ { M '}_{ p }  } \right]^{J_{ A }}_{M_{ A }} \label{Phi_prime}
    \end{eqnarray}
    Antisymmetrizers have been suppressed in Eq. (\ref{expansion_second_sum}) except for $H_{Tp}$ due to Eqs.(\ref{HT_action},\ref{Hp_action}).
    The term involving Kronecker delta vanishes in Eq. (\ref{expansion_second_sum}) as $N \neq N'$ and $H_{Tp}$ coupling can be neglected therein as $N' > N_{ max }$.
   
    As a consequence, the second sum of Eq.(\ref{matrix_elmts_h_k_decomposed}) is equal to zero and its third sum is treated identically for symmetry reasons. In the last sum, antisymmetrizers of Eq. (\ref{expansion_second_sum}) can be suppressed as $N > N_{ max }$ and $N' > N_{ max }$. 

    Therefore, we have:
    \begin{eqnarray}
      &&H_{ cc' }^{ J_{ A } M_{ A }} \left( N , N' \right)  \nonumber \\ 
      &=& \bra{\Phi_{A}} \hat{ H } \ket{\Phi'_{A}} \nonumber \\
      &=& \bra{\Phi_{A}} \hat{ H }_{ T } + \hat{ H }_{ p }  \ket{\Phi'_{A}} \nonumber \\
      &=& \bra{\Phi_{A}} \hat{ H }_{ T } + \hat{ H }_{\text{int}} + \hat{ H }_{CM} \ket{\Phi'_{A}} \nonumber \\
      &=& \left( E_{ T } + E_{ int } + E_{ CM } \right) { \delta }_{ cc' } \delta_{ N N'}
      \label{matrix_elmts_h_high_energy}
    \end{eqnarray}
    
    Consequently, we can express the matrix elements $H_{ cc' }^{ J_{ A } M_{ A }} (R , R')$ as: 
    \begin{eqnarray}
      && H_{ cc' }^{ J_{ A } M_{ A }} (R , R') \nonumber \\ 
      & = &  \sum_{ \stackrel{N  \leq N_{ max }} { N' \leq N_{ max }} } H_{ cc' }^{ J_{ A } M_{ A }} (N , N') \frac{ U_{N} (R) }{ R } \frac{ U_{ N'} (R') }{ R' } \nonumber \\
      & + & \!\!  { \delta }_{ cc' } \!\!\!\!\!\!\! \sum_{ \stackrel{N  > N_{ max }}{N' > N_{max}} } \!\!\!\!\!\! \left( E_{ T } +  E_{ int } +  E_{ CM } \right) \frac{ U_{N} (R) }{ R } \frac{ U_{N'} (R') }{ R' }
      \label{matrix_elmts_h_2}
    \end{eqnarray}
    
    The sums in Eq. (\ref{matrix_elmts_h_2}) involving $N > N_{ max }$ and $N' > N_{ max }$ can be written as:
    \begin{eqnarray}
      && \sum_{ \stackrel{N  > N_{ max }}{N' > N_{max}} } \left( E_{ T } + E_{ int } + E_{ CM } \right) \frac{ U_{N} (R) }{ R } \frac{ U_{N'} (R') }{ R' } \nonumber \\
      & = &  \sum_{ N N' } \left( E_{ T } +  E_{ int }  +  E_{ CM } \right) \frac{ U_{N} (R) }{ R } \frac{ U_{N'} (R') }{ R' } \nonumber \\
      & - &  \sum_{ \stackrel{N \leq N_{ max }}{ N' \leq N_{max} }} \left( E_{ T } +  E_{ int }  +  E_{ CM } \right) \frac{ U_{N} (R) }{ R } \frac{ U_{N '} (R') }{ R' } 
      \label{last_sum}
    \end{eqnarray}
    where the the first term in Eq. (\ref{last_sum}) can be expressed with Dirac delta's due to completeness properties of $U_{ N}(R)$ states:
    \begin{align}
      & \sum_{ N N' } \left( E_{ T } +  E_{ int } + E_{CM} \right)  \frac{ U_{N} (R) }{ R } \frac{ U_{N} (R') }{ R' } \nonumber \\
      & = \left( E_{ T } +  E_{ int } + \hat{T}_{CM} \right) \frac{ \delta (R - R') }{ R R' } { \delta }_{ cc' } + { U^{ L }_{CM}  } (R , R') { \delta }_{ cc' } 
      \label{sum_with_completeness_relation}
    \end{align}
    where $\hat{T}_{CM}$ and $U^{ L }_{CM} (R , R')$ (for a fixed orbital momentum ${ L }$) stand for the center-of-mass kinetic and potential part of Eq. (\ref{HCM}) respectively.
    Hence, we can finally write the expression of $H_{ cc' }^{ J_{ A } M_{ A }} (R , R')$:
    \begin{eqnarray}
      && H_{ cc' }^{ J_{ A } M_{ A }} (R , R') \nonumber \\
      &=& \frac{ { \hbar }^{ 2 }}{ 2 M_{p}} \left( -\frac{ { d }^{ 2 }}{ d R^{ 2 }} + \frac{L(L + 1)}{R^2} \right) \frac{ \delta (R - R') }{ R R' } { \delta }_{ cc' }   \nonumber \\
      &+& \left( E_{ T } + E_{ int } \right) \frac{ \delta (R - R') }{ R R' } { \delta }_{ cc' }  \nonumber \\
      &+& { U_{CM}^{ L }} (R , R') ~ { \delta }_{ cc' } + { \tilde{ V }}_{ cc' }^{ J_{ A } M_{ A }} (R , R')
      \label{hamiltonian_matrix_elmts}
    \end{eqnarray}
    where $\tilde{ V }_{ cc' }^{ J_{ A } M_{ A }}$ includes the remaining short-range potential terms of the Hamiltonial kernels, 
    i.e.~the first sum of Eq. (\ref{matrix_elmts_h_k_decomposed},\ref{matrix_elmts_h_2}) and the last sum of Eq. (\ref{last_sum}).
    $N_{ cc' }^{ J_{ A } M_{ A }} (N , N')$ many-body matrix elements are calculated using the Slater determinant expansion of the cluster wave functions.
    The treatment of the non-orthogonality of channels is the same as in the one-nucleon projectile case \cite{PRC_GSM_CC_Yannen}.

    {
      As mentioned earlier, it is necessary to use a basis of harmonic oscillator states to calculate coupled channel-potentials. 
      For this, the $\ket{N ~ L}$ eigenstates of $\hat{ H }_{CM}$ just have to be expanded in a HO basis by HO states $\ket{N ~ L}^{\text{HO}}$ to calculate $\tilde{ V }_{ cc' }^{ J_{ A } M_{ A }}$.
      Equations(\ref{HO_sd},\ref{HO_Berggren_sd}) can then be used to apply shell model formulas to calculate $H_{ cc' }^{ J_{ A } M_{ A }} (N , N')$ and $N_{ cc' }^{ J_{ A } M_{ A }} (N , N')$ many-body matrix elements.
      Note that the antisymmetry of channels (see Eq. (\ref{channel})) is exactly taken into account through the expansion of many-body targets and projectiles with Slater determinants.
    }

  }

  \subsection{Berggren basis expansion methods to solve coupled-channel equations}
  \label{numerical_resolution}
  
  In this section, we present the different numerical techniques used to calculate the $A$-body scattering states $\ket{\Psi_{ M_{ A }}^{ J_{ A }}}$.
  Hence, we shall present how to solve numerically the radial wave functions $u_{ c } (r)$ based on a Berggren basis expansion of the Green function ${ { (H - E) }^{ -1 } }$.
  Indeed, in previous papers involving GSM-CC \cite{PRC_GSM_CC_Yannen,PRC_GSM_CC_Kevin}, a direct integration method had been used.
  However, direct integration has been noticed to become unstable when channel coupling is very strong, and to overcome this problem we have developped a new method whose implementation takes advantage of the Berggren basis complex energies.
  It is based on the representation of $H$ with the center-of-mass Berggren basis generated by Eq. (\ref{HCM}), which replaces integro-differential equations by a linear matrix problem.
  Note that this method can also be used in the one-nucleon projectile case, so that we will include this case in the following discussion.

  For this, we start from the $A$-body scattering state $\ket{\Psi_{ M_{ A }}^{ J_{ A }}}$, of energy $E$, which is a solution of the Schr\"odinger equation:
  \begin{equation}
    \hat{ H } \ket{\Psi_{ M_{ A }}^{ J_{ A }} } = E \ket{ \Psi_{ M_{ A }}^{ J_{ A }}  } \label{W_equation} \; .
  \end{equation}
  
  The $A$-body scattering state is decomposed on a channel basis as:

  \begin{equation}
    \ket{\Psi_{ M_{ A }}^{ J_{ A }} } = \sum_{c} \int_{0}^{\infty} \frac{ { u }_{ c } (r) }{ r } { r }^{ 2 } \ket{(c,r) } dr \; .
    \label{}
  \end{equation}

  Here, $u_{ c }(r)$ are radial wave function associated to the channel $c$.
  Note that the radial $r$ distance stands for either the distance between the nucleon or the compound projectile and the target.

  Due to the antisymmetry of the target-projectile composite, the channel functions ${ { u }_{ c }(r) }$ are not orthogonal. In order to consider a matrix representation of channels, one uses the standard method described in Ref.\cite{PRC_GSM_CC_Yannen} with the notation of this paper, and which consists in using ${ { w }_{ c }(r) }$ orthogonal channel functions. Those functions are solutions of the coupled-channel equations when considering the matrix ${ { \hat{ N } }^{ -\frac{ 1 }{ 2 } } \hat{ H } { \hat{ N } }^{ -\frac{ 1 }{ 2 } } }$, where matrix elements of ${ \hat{ H } }$ and ${ \hat{ N } }$ are given by ${ { \hat{ H } }_{ cc' }(R,R') }$ and ${ { \hat{ N } }_{ cc' }(R,R') }$ respectively. The ${ { u }_{ c }(r) }$ channel functions are recovered from ${ { w }_{ c }(r) }$ channel functions at the end of the calculation using ${ { \hat{ N } }^{ -\frac{ 1 }{ 2 } } }$ as well \cite{PRC_GSM_CC_Yannen}. In order not to complicate notations, we will implicitly consider in the rest of this section that the $c$ channels are the orthogonalized channels associated to $w_c(r)$ functions, even though they are in principle  linear combinations of the initial channels defining $u_c(r)$ channel functions.

  In order to use the Berggren basis to invert $H - E$, we introduce an approximate Hamiltonian $\hat{ H }^{(0)}$ and its eigenvector $\ket{\Psi^{(0)}}$:
  \begin{align}
    { \hat{ H }}^{(0)} & = \hat{ t } + \hat{U}_{basis} \mbox{ (nucleon)} \nonumber \\
    & = { \hat{ T }}_{CM} + \hat{U}_{CM} \mbox{ (cluster)} \label{H0_definition} \\
    { \hat{ H }}^{(0)} \ket{{\Psi}^{(0)}} & = E \ket{{\Psi}^{(0)}} \label{W0_definition} \; .
  \end{align}
  where $\hat{U}_{basis}$ has the same form as $\hat{U}_{ basis }^{ T }$.

  More specifically $\hat{ H }^{ (0) }$ is a diagonal matrix (no coupling with other channels) and only one non-zero value (only the entrance channel $c_{ 0 }$ is active). 
  Equation (\ref{W0_definition}) is straightforward to solve as $\hat{ H }^{(0)}$ leads to a one-dimensional differential equation.
  
  Let us then separate $\hat{ H }$ and $\ket{\Psi_{M_{ A }}^{ J_{ A }}}$ in two parts, involving $\hat{ H }^{(0)}$ and $\ket{\Psi^{(0)}}$ and a rest part:
  \begin{align}
    \hat{ H } & = { \hat{ H }}^{(0)} + \hat{ H }_{rest} \label{H_separation} \\ 
    \ket{\Psi_{ M_{ A }}^{ J_{ A }} } & = \ket{{ \Psi }^{(0)}} + \ket{\Psi_{rest}} \label{W_separation} \; .
  \end{align}
  
  Using Eqs.(\ref{W_equation},\ref{W0_definition},\ref{H_separation},\ref{W_separation}), one obtains:
  \begin{align}
    (\hat{ H } - E) \ket{\Psi_{rest}} & = \ket{S} \label{Wrest_equation} \\ 
    \ket{S} & = - \hat{ H }_{rest} \ket{{ \Psi }^{(0)}} \label{source_definition} \; ,
  \end{align}
  where the source term $\ket{S}$ has been introduced.
  Since all the long range contribution is contained within ${ { \hat{ H } }^{ (0) } }$, $\hat{ H }_{rest}$ is of finite range, $S(r) \rightarrow 0$ when $r \rightarrow +\infty$ from Eq. (\ref{source_definition}). 
  Hence, $\ket{S}$ can be expanded in the Berggren basis generated by $\hat{ H }^{(0)}$, so that Eq. (\ref{Wrest_equation}) becomes a linear system in this representation:
  \begin{align}
    (\Psi_{rest})_{n,c} & = \braket{n,c | \Psi_{rest}} \label{W_vector_definition} \\
    (M_E)_{n,c~n',c'} & = \braket{n',c' | \hat{ H } - E | n,c} \label{matrix_definition} \\
    (S)_{n,c} & = \braket{n,c | S} \label{S_vector_definition} \\
    M_E ~ \Psi_{rest} & = S \label{W_linear_system}
  \end{align}
  where $\ket{n,c}$ is a Berggren basis state of index $n$ of the channel $c$.
  
  The fundamental problem of Eq. (\ref{W_linear_system}) is the non-invertible character of $M_E$ on the real axis, as $H$ possesses a scattering eigenstate of energy $E$ therein.
  The standard solution to this problem is to replace $E$ by $E + i \epsilon$, with $\epsilon \rightarrow 0^+$, 
  so as to make the considered linear system invertible, on the one hand, and to impose an outgoing wave function character of $u_c(r)$ on all outgoing channels, on the other hand.
  As  $w_c(r)$ and $u_c(r)$ channel functions become equal asymptotically due to the disappearance of antisymmetry between target and projectile at large distance, this also provides with outgoing wave function behavior of $u_c(r)$ channel functions.
  
  However, this method becomes unstable for small $\epsilon$ and demands to carefully monitor the limiting process. 
  In order to avoid this problem, the contour in the complex plane defining the $\ket{n,c}$ Berggren basis states is chosen so that the energy of basis states always has a non-zero imaginary part.
  Consequently, $M_E$ is invertible using this contour, so that Eq. (\ref{W_linear_system}) is numerically solvable without introduction of a regularization parameter.
  The outgoing wave function character of $\Psi_{rest}$ is also guaranteed by the finite norm of $\Psi_{rest}$ in a Berggren basis representation. 
  Indeed, as $||\Psi_{rest}||$ is finite, one can deduce from the Parseval equality extended to Berggren bases that $\ket{\Psi_{rest}}$ is a localized state when complex rotation is applied,
  i.e.~$\Psi_{rest}(z) \rightarrow 0$ when $z \rightarrow +\infty$, with $z = r + (R - r) e^{i \theta}$, with $R$ a radius outside the nuclear zone and $0 < \theta < \pi/2$ being properly chosen. 
  This implies that $\ket{\Psi_{rest}}$ has an outgoing character in all channels. 
  
  Having calculated $\ket{\Psi_{rest}}$ in a Berggren basis, its calculation in coordinate space is straightforward. 
  In cases when the equivalent potential method \cite{Michel_equivalent_potential} is numerically stable, it has been checked numerically that both Berggren basis expansion and direct integration methods provide with the same $\ket{\Psi_{ M_{ A }}^{ J_{ A }}}$ solution. The Berggren basis is also useful to determine resonant states of the coupled-channel Hamiltonian $\hat{ H }$, where the coupled-channel equations become a diagonalization matrix problem.
  
  Equation(\ref{W_linear_system}) also leads to an additional numerical advantage when many energies have to be considered. Indeed, the spectrum of $\hat{ H } - E$ is the same for all energies.
  Consequently, it is sufficient to calculate a convenient representation of $\hat{H}$ only once,
  which can then be reused to solve the linear system of Eq. (\ref{W_linear_system}) where different energies are considered.
  In practice, this replaces many dense linear systems to solve, e.g.~with the standard LU decomposition \cite{LU_decomposition},
  whose numerical cost is about $N^3/3$ per energy, where $N$ is the dimension of $M_E$, 
  by a diagonal or tridiagonal linear system to solve and two matrix multiplications, whose numerical cost is about $2 N^2$ per energy.

\section{Model space and Hamiltonian}
\label{sec3}

GSM-CC with cluster projectiles will be applied to the $^4$He(d,d) elastic scattering reactions and asymptotic normalization coefficients of the $^6$Li wave function. In the following application, only deuteron reaction channels will be used in the decomposition of low-energy states of $^6$Li.

The internal structure of deuteron is calculated using the N$^3$LO interaction \cite{PRC_N3LO}, fitted on phase shifts properties of proton-neutron elastic scattering reactions.
The used realistic interaction is firstly diagonalized with a two-body intrinsic Berggren basis generated by a Woods-Saxon potential in order to generate intrinsic deuteron states.
Its diffuseness, radius, central and spin-orbit strengths are 0.65 fm, 1.5 fm, 40 MeV and 7.5 MeV, respectively. 

Berggren contours consist of two real segments, defined by the origin of the real $k$-axis and the $k$-points equal to 0.2 fm$^{-1}$ and 2 fm$^{-1}$. They are discretized with 2 and 8 points, respectively. It has been checked that having a finer discretization does not change numerical results significantly.
One then obtains a deuteron ground state energy of -2.061 MeV, close to the experimental value of -2.224 MeV.
As stated in Sec. \ref{sec2.1}, the Berggren eigenstates obtained after diagonalization are projected on a basis of laboratory HO states afterwards.

The Berggren basis of center-of-mass cluster states, used to calculate resonant and scattering states of d + $^4$He ($^6$Li) aggregate (see Sec. \ref{sec2.2}), is generated by proton/neutron Woods-Saxon potentials using Eqs. (\ref{UCM_central},\ref{UCM_so}). Different $S$, $P$, $D$, $F$ and $G$ partial waves bear 3, 2, 1, 1 and 0 bound pole states, respectively, which are included along with the respective contours. The contours consist of three segments, defined by the origin of the $K_{CM}$ complex plane and the complex points $K_{CM}$: 0.2-i0.05 fm$^{-1}$, 1.0-i0.05 fm$^{-1}$ and 2 fm$^{-1}$. Each segment is discretized with 15 points, so that each contour possesses 45 points.
All unbound pole states lie below the Berggren basis contours, so that they do not belong to the considered Berggren bases.

It has been checked that the most important intrinsic deuteron wave functions are those bearing $\left( J^{\pi} \right)_{\text{int}} = 1^+$, $3^+$, so that we consider only those deuteron intrinsic wave functions. 
We build relative Berggren bases states with $\left( J^{\pi} \right)_{\text{int}} = 1^+$, $3^+$, so that the relative scattering states present therein mimic deuteron break-up, similarly to the deuteron pseudo-states of Ref. \cite{PRL_Hupin} in NCSMC calculation. Center-of-mass parts of deuteron projectiles bear $L_{CM} \leq 4$ for $\left( J^{\pi} \right)_{\text{int}} = 1^+$ and $L_{CM} = 0$ for $\left( J^{\pi} \right)_{\text{int}} = 3^+$. The total angular momentum of deuteron projectiles has been chosen to verify $J_{d} \leq 3$, as in Ref. \cite{PRL_Hupin}.

We use $^4$He core with two valence nucleons to describe $^6$Li wave functions. All partial waves up to $\ell = 4$ are included.
The $^4$He core is mimicked by a Wood-Saxon potential, fitted on phase shifts of elastic scattering reactions involving a neutron and proton on a $^4$He target (see Table \ref{WS_parameters}, issued from Ref.\cite{PRC_FHT_Yannen}).
\begin{table}[htb]
\caption{\label{WS_parameters} 
Parameters of the proton and neutron Woods-Saxon potentials fitted from $^4$He-nucleon phase shifts. From top to bottom: central potential depth, spin-orbit potential depth, radius, diffuseness and charge radius. See Ref. \cite{PRC_FHT_Yannen} for details concerning its derivation.}
\begin{ruledtabular}
\begin{tabular}{lcc}
Parameter & Neutrons & Protons \\ \hline\\
$V_0$  (MeV) &  41.9   & 44.4\\
$V_{\ell s}$ (MeV\, fm$^2$) &  7.2  &7.2 \\
$R_0$ (fm) &  2.15 &  2.06 \\
$a$ (fm)  & 0.63 &  0.64 \\
$R_{\rm ch}$ (fm)  &--  & 1.681
\end{tabular}
\end{ruledtabular}
\end{table}
The nucleon-nucleon interaction is that of FHT type \cite{1979Furutani,1980Furutani}, which has been recently fitted \cite{PRC_FHT_Yannen} for light nuclei bearing a $^4$He core. It reads:
\begin{equation}
V_{FHT} = V_c + V_{LS} + V_{T} \label{V_FHT}
\end{equation}
where $V_c$ , $V_{LS}$, $V_{T}$ represent its central, spin-orbit and tensor part, respectively. There is no Coulomb part as we have only one valence proton in $^6$Li wave functions.
The different components $V_{FHT}$ in Eq. (\ref{V_FHT}) read \cite{PRC_FHT_Yannen}:
\begin{eqnarray}
V_c(r) &=&  \sum_{n=1}^3  \, V_c^n   \,    \left( W_c^n + B_c^n P_{\sigma} - H_c^n  P_{\tau} \right. \nonumber \\
&& \qquad\qquad\qquad\quad \;\:\;\left.- M_c^n P_{\sigma}P_{\tau} \right)\: e^{-\beta_c^n r^2}  \label{FHT1} \\
V_{LS}(r) &=& \vec{L}\cdot \vec{S}\;\, \sum_{n=1}^2 \, V_{LS}^n  \, \left( W_{LS}^n - H_{LS}^n  P_{\tau}  \right) \, e^{-\beta_{LS}^n r^2}   \label{FHT2} \\
V_{T}(r) &=&  S_{ij}\:\sum_{n=1}^3 \,  V_{T}^n \, \left( W_{T}^n  - H_{T}^n  P_{\tau} \right) \, r^2 e^{-\beta_{T}^n r^2},   \label{FHT3}
\end{eqnarray}
where $r$ is the distance between the nucleons $i$ and $j$, 
$\vec{L}$ is the relative orbital angular momentum,  
$\vec{S}=(\vec{\sigma}_i+\vec{\sigma}_j) / 2$ is the total spin of the two nucleons, 
$S_{ij}=3 (\vec{\sigma}_i \cdot \hat{r}) (\vec{\sigma}_j \cdot \hat{r})   - \vec{\sigma}_i \cdot \vec{\sigma}_j$ is the tensor operator, 
$P_{\sigma}$ and  $P_{\tau}$ are spin and isospin exchange operators, respectively,
$V_c^n$, $n \in \{1,2,3\}$ and $V_{LS}^n$, $V_{T}^n$, $n \in \{1,2\}$ are parameters fitting the central, spin-orbit and tensor, respectively, while other parameters are fixed \cite{1979Furutani}.
Following Ref.\cite{PRC_FHT_Yannen}, we rewrite $V_{FHT}$ in terms of its spin and isospin dependence:
\begin{eqnarray}
V_c(r) &=&  V_c^{11} \, f_c^{11}(r)  \Pi_{11} \: + \:  V_c^{10} \, f_c^{10}(r) \Pi_{10} \nonumber  \\
&+& V_c^{00} \, f_c^{00}(r) \Pi_{00} \: + \:  V_c^{01} \, f_c^{01}(r) \Pi_{01},  \label{inter1}  \\
V_{LS}(r) &=& (\vec{L}\cdot \vec{S}) \,V_{LS}^{11} \, f_{LS}^{11}(r) \Pi_{11}, \label{inter2}  \\
V_{T}(r) &=& S_{ij} \left[V_{T}^{11}f_{T}^{11}(r) \Pi_{11} + V_{T}^{10} f_{T}^{10}(r) \Pi_{10}\right],\label{inter3}
\end{eqnarray}
where $\Pi_{ST}$ are projectors on spin and isospin \cite{1974DeShalit,RingSchuck} and $f_c^{ST}(r)$, $f_{LS}^{ST}(r)$ and  $f_{T}^{ST}(r)$ functions are straightforward to evaluate from Eqs.(\ref{FHT1},\ref{FHT2},\ref{FHT3}). 
Matrix elements of the Hamiltonian are calculated in the model space consisting of all proton and neutron HO states having $\ell \leq 4$ and $n \leq 5$. 
The use of Berggren basis at this level is not necessary as the Slater determinants used therein only generate the GSM-CC Hamiltonian interaction ${ { \tilde{ V } }_{ cc' }^{ { J }_{ A } { M }_{ A } }(R,R') }$, which is finite ranged (see Sec. \ref{sec2.3} and Eq.(\ref{hamiltonian_matrix_elmts})).
Note that the use of HO states does not hamper the asymptotes of the loosely bound and resonance states of $^6$Li. Indeed, HO states are used only to generate the finite range part of the GSM-CC Hamiltonian, whereas the eigenstates of $^6$Li are expanded with Berggren basis states. Consequently, the density of $^6$Li eigenstates slowly decreases or increases exponentially in modulus, respectively, independently of the Gaussian fall-off of HO states. Convergence for Hamiltonian representation is typically obtained with 5-10 HO states per partial wave \cite{PRC_isospin_mixing}.

The statistical properties of the FHT interaction parameters for $p$-shell nuclei have been analyzed in Ref.\cite{PRC_FHT_Yannen}. It has been noticed that they bear a sizable statistical error.
Consequently, one can modify the FHT interaction parameters within the bounds of calculated statistical errors without in principle changing the interaction.

The $T=1$ interaction part is negligible both in the $^4$He(d,d) reaction and in the $T=0$ spectrum of $^6$Li. The dependence of energies on $V_c^{00}$ is very weak, so that we only consider $V_c^{10}$ and $V_{T}^{10}$ when fitting the FHT interaction. Other parameters of the FHT interaction remain the same as in Ref. \cite{PRC_FHT_Yannen}.
It has been found that in this way only ground state energy of $^6$Li can be fitted satisfactorily whereas all other $T=0$ resonances are displaced significantly with respect to their experimental energy centroids. Moreover, the asymptotic normalization coefficients $^3S_1$ and $^3D_1$ of the $^6$Li wave function in d + $^4$He configuration are too small as compared to the reported experimental values \cite{ANCs_exp_1,ANCs_exp_2,ANCs_exp_3} and equal 1.918 and -0.051, respectively. On the contrary, the ratio of $^3D_1$/$^3S_1=-0.0265$ agrees well with the experimental value -0.025(6)(10) \cite{ANCs_exp_1,ANCs_exp_2,ANCs_exp_3}.

In the following, two different strategies of fitting the FHT interaction have been employed.
The first strategy (FHT(E)) comes from a fit of both the $T=0$ spectrum of $^6$Li and the value of $^3D_1$ asymptotic normalization coefficient.  
The second strategy (FHT(ANC)) corresponds to fitting the resonant $T=0$ spectrum of $^6$Li, along with the $^3S_1$ and $^3D_1$ asymptotic normalization coefficients, leaving the ground state energy of $^6$Li out of the fit (see Table \ref{ANCs}). 
In both strategies, in addition to the modification of  $V_c^{10}$ and $V_{T}^{10}$, the $T=0$ matrix elements bearing $J^\pi = 2^+, 3^+$ are multiplied by a small factor $c(J^\pi)$ depending on $J^\pi$, to reproduce energies of the $2^+$ and $3^+$ resonant $T=0$ states of $^6$Li (see Table \ref{FHT_parameters}). 
\begin{table}[htb]
\caption{\label{ANCs}
$^3S_1$ and $^3D_1$ asymptotic normalization coefficients and their ratio calculated in GSM-CC is compared with the experimental values. 
In FHT(ANC), both $^3S_1$ and $^3D_1$ asymptotic normalization coefficients have been fitted to reproduce the values reported in Ref. \cite{ANCs_exp_1,ANCs_exp_2,ANCs_exp_3}. In FHT(E), only the $^3D_1$ asymptotic normalization coefficient has been fitted.}
\begin{ruledtabular}
\begin{tabular}{lcccc}
ANC          & FHT(E)  & FHT(ANC) & Exp \cite{ANCs_exp_1,ANCs_exp_2,ANCs_exp_3} & Exp \cite{blokh93,veal98} \\ \hline 
$^3S_1$ (fm$^{-1/2}$)      & 1.707     & 2.950       & 2.91(9)  & 2.93(15) \\
$^3D_1$ (fm$^{-1/2}$)     & -0.0788   & -0.077      & -0.077(18)  & - \\
$^3D_1/^3S_1$ & -0.0462   & -0.0261     & -0.025(6)(10)  & 0.0003(9) \\
\end{tabular}
\end{ruledtabular}
\end{table}
\begin{table}[htb]
\caption{\label{FHT_parameters} 
Parameters of the FHT(E) and FHT(ANC) interactions are compared with the original FHT parameters with their statistical uncertainties reported in \cite{PRC_FHT_Yannen} for $p$-shell nuclei. The $T=1$ and $V_c^{00}$ parameters are written as well, even though they are not fitted. }
\begin{ruledtabular}
\begin{tabular}{lccc}
Parameter   & FHT \cite{PRC_FHT_Yannen}     & FHT(E)  & FHT(ANC)  \\ \hline 
$V_c^{11}$   & -3.2 (220) & -3.2         & -3.2   \\
$V_c^{10}$   & -5.1 (10) & -5.41        & -6.675 \\
$V_c^{00}$   & -21.3 (66) & -21.3        & -21.3  \\
$V_c^{01}$   & -5.6 (5) & -5.6         & -5.6   \\
$V_{LS}^{11}$ & -540 (1240) & -540         & -540   \\
$V_{T}^{11}$  & -12.1 (795) & -12.1        & -12.1  \\
$V_{T}^{10}$  & -14.2 (71) & -10.5        & -4.1   \\
$c(2^+)$    & - & 1.16         & 1.068  \\
$c(3^+)$    & - & 1.1314       & 0.9802
\end{tabular}
\end{ruledtabular}
\end{table}
In this way, we obtain two effective interactions FHT(E) and FHT(ANC) which will be tested in the following (see Table \ref{FHT_parameters}). 

The use of two different interactions to deal with the structure of $^6$Li and the elastic scattering of deuteron on $\alpha$-particle is necessary as we have two different pictures in our model: 
that of a deuteron far from the $^4$He target before and after the reaction occurs, where deuteron properties are prominent, and that of a $^6$Li composite during the reaction.
As the FHT interaction is defined from $^6$Li properties, it cannot grasp the structure of deuteron at large distances.

Conversely, the N$^3$LO interaction cannot be used with a core. 
Moreover, as the N$^3$LO interaction enters only the deuteron projectile basis construction, it is not explicitly present in the Hamiltonian, but just insures that the deuteron projectile has correct both the asymptotic behavior and the binding energy. This also implies that the use of both laboratory and COSM coordinates is consistent therein, as they coincide asymptotically.

It has been indeed checked that the use of AV8 \cite{PRL_AV8} and CD-Bonn \cite{PRC_CD_Bonn} interactions to generate deuteron projectiles leads to a very small change in energies, asymptotic normalization coefficients and cross sections. As a consequence, the use of both realistic interaction for projectiles and effective Hamiltonian for composites induces no problem in our framework.

\section{Results}
\label{sec4}

\subsection{$T=0$ low-energy spectrum}
\label{sec4.1}

\begin{figure}[htb]
\includegraphics[width=9.5cm]{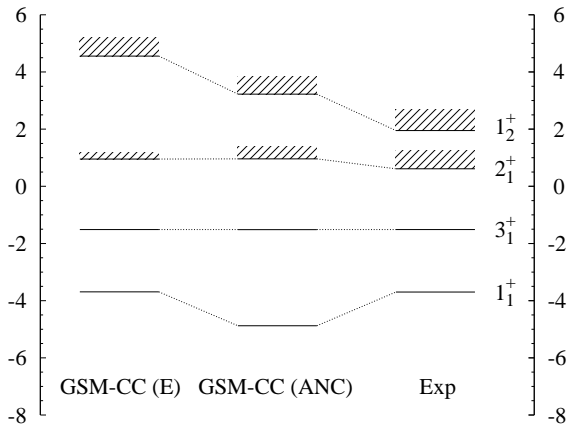}
\caption{Energies (given with respect to ${ {  }^{ 4 } }$He core) and widths (in MeV) of the $T=0$ spectrum of $^6$Li calculated in GSM-CC approach using FHT(E) and FHT(ANC) interactions and denoted as GSM-CC(E) and GSM-CC(ANC), respectively,  are compared to evaluated energy centroids and widths of Ref. \cite{ANCs_exp_1}. }
\label{spectra}
\end{figure}

The energies and widths of the $T=0$ spectrum of $^6$Li calculated using FHT(E) and FHT(ANC) sets of parameters are compared in Fig. \ref{spectra} with the energy centroids and the widths determined in Ref. \cite{ANCs_exp_1}. 
To obtain the spectra from GSM-CC calculations, the Hamiltonian is represented in the coupled-channel representation and then diagonalized.

As seen in Fig. \ref{spectra}, the ground state of $^6$Li in the GSM-CC(ANC) calculation is overbound in order to reproduce results of Refs. \cite{ANCs_exp_1,ANCs_exp_2,ANCs_exp_3} for both $^3S_1$ and $^3D_1$ asymptotic normalization coefficients (see Table \ref{ANCs}). However, as the asymptotic normalization coefficients are well reproduced, one can expect that this interaction also provides a good reproduction of the cross sections.

Widths of resonance states are described qualitatively in the GSM-CC calculations. The width of the $3^+_1$ state, not visible in the figure, is about 4 keV for both FHT(E) and FHT(ANC) parametrizations, whereas  the experimental width is 24 keV \cite{ANCs_exp_1}. The width of the $2^+_1$ state is 500 keV and 840 keV for the FHT(E) and FHT(ANC) interactions, respectively, while the reported experimental value is 1.3 MeV \cite{ANCs_exp_1}. The resonance energy is about 300 keV higher than found in $R$-matrix analyses of the experimental data \cite{ANCs_exp_1}. This deliberate underbinding is necessary for $^3D_2$ phase shifts to be optimally described (see the discussion around Fig. \ref{phase_shifts}). 
The $1^+_2$ state is underbound, by 2.6 MeV and 1.3 MeV in GSM-CC(E) and GSM-CC(ANC) calculation, respectively.  The calculated width for this state is 1.293 MeV (GSM-CC(E)) and 1.226 MeV (GSM-CC(ANC)), as compared to 1.5 MeV reported in Ref. \cite{ANCs_exp_1}.

\subsection{Energy dependence of phase shifts}
\label{sec4.2}

\begin{center}
\begin{figure}[htb]
\includegraphics[width=9.5cm]{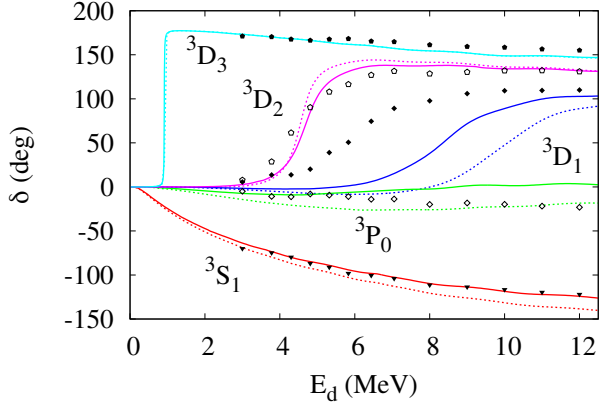} 
\vskip -0.5truecm
\caption{(Color online) Phase shifts of the $^4$He(d,d) elastic scattering reaction calculated using the FHT(E) interaction (dotted lines) and the FHT(ANC) interaction (solid lines) are
  compared to results of the $R$-matrix analyses of experimental data (symbols) \cite{NPA_phase_shifts_exp_1,NPA_phase_shifts_exp_2}. $E_d$ is the kinetic energy of the incoming deuteron and is expressed in the laboratory frame.}
\label{phase_shifts}
\end{figure}
\end{center}

Phase shifts of the $^4$He(d,d) elastic scattering reaction are represented in Fig. \ref{phase_shifts}. One can see that the phase shifts extracted from $R$-matrix analyses of data \cite{NPA_phase_shifts_exp_1,NPA_phase_shifts_exp_2} are well described qualitatively using both FHT(E) and FHT(ANC) interactions, except for the $^3D_1$ phase shifts, as the $1^+_2$ state lies too high in energy for both interactions. 

The FHT(ANC) interaction provides the best reproduction of phase shifts, especially for the $^3S_1$ and $^3P_0$ channels. 
The $^3D_1$ phase shifts are also comparatively closer to the data in GSM-CC(ANC) than in GSM-CC(E). The $^3D_2$ phase shift increases too rapidly in the GSM-CC(E), whereas it is closer to experiment in the GSM-CC(ANC) calculation. In both GSM-CC(E) and GSM-CC(ANC) it was necessary to have the energy of $2^+_1$ state (the $2^+_1$ pole of the $S$-matrix) about 300 keV above the value reported in Ref. \cite{ANCs_exp_1} to obtain the $^3D_2$ phase shift well centered on the experimental resonance. This is directly related to the large width of the $2^+_1$ state, implying that the many-body $S$-matrix poles provided by GSM-CC calculation and the resonance structures seen in reaction observables are not equivalent. Consequently, reproducing resonant states energies in structure calculation does not necessarily guarantee a good reproduction of the experimental cross sections. Nevertheless, it is clear that for all channels the FHT(ANC) parametrization provides the phase shifts that are closer to those extracted from $R$-matrix analyses than the phase shifts obtained for the FHT(E) interaction. Therefore, one can expect a better reproduction of the experimental cross sections using the FHT(ANC) interaction than the FHT(E) interaction.

\subsection{Differential cross sections}
\label{sec4.3}

\begin{figure}[htbp]
\includegraphics[width=8.6cm]{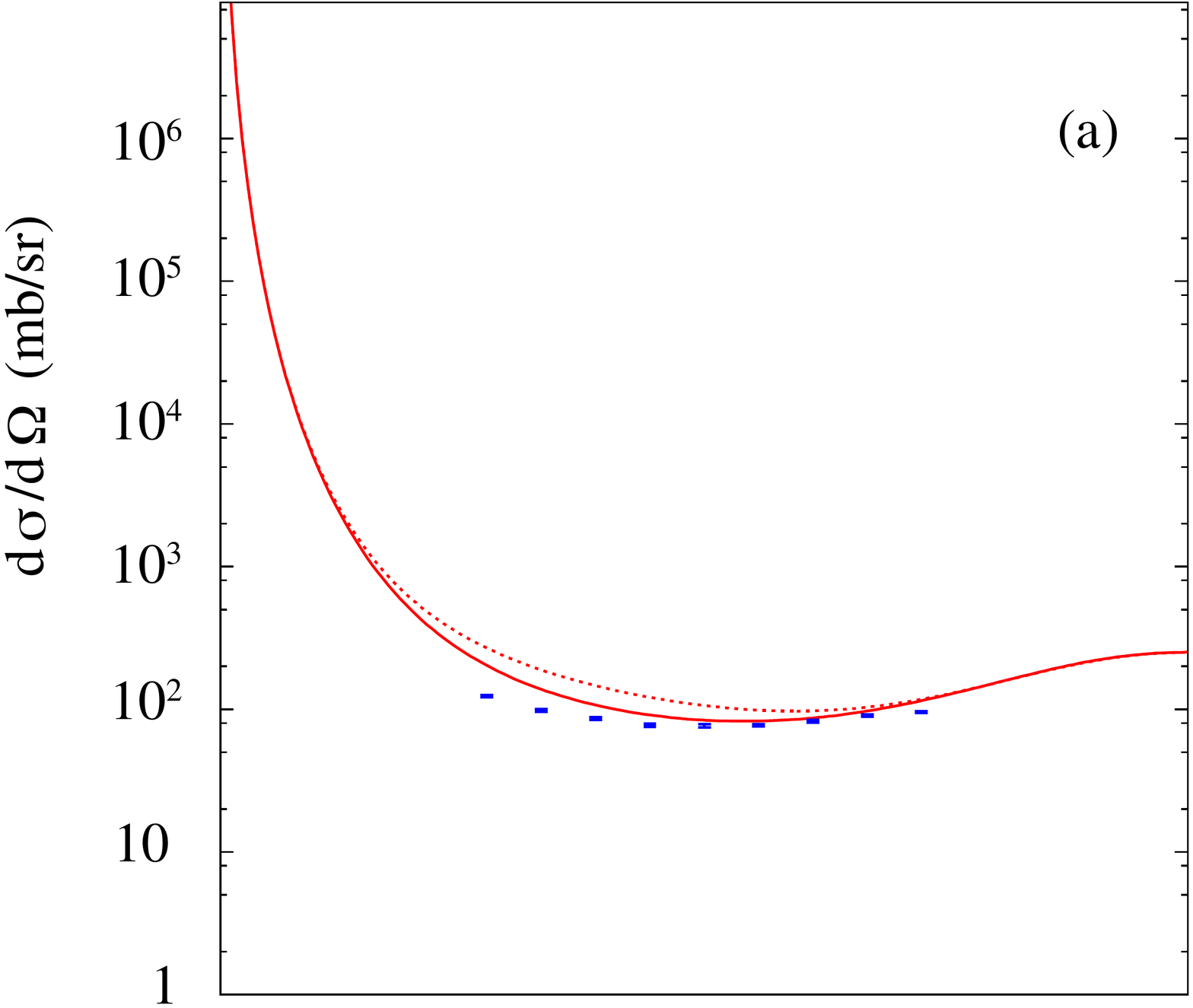} \\
\vskip -2truecm
\includegraphics[width=8.65cm]{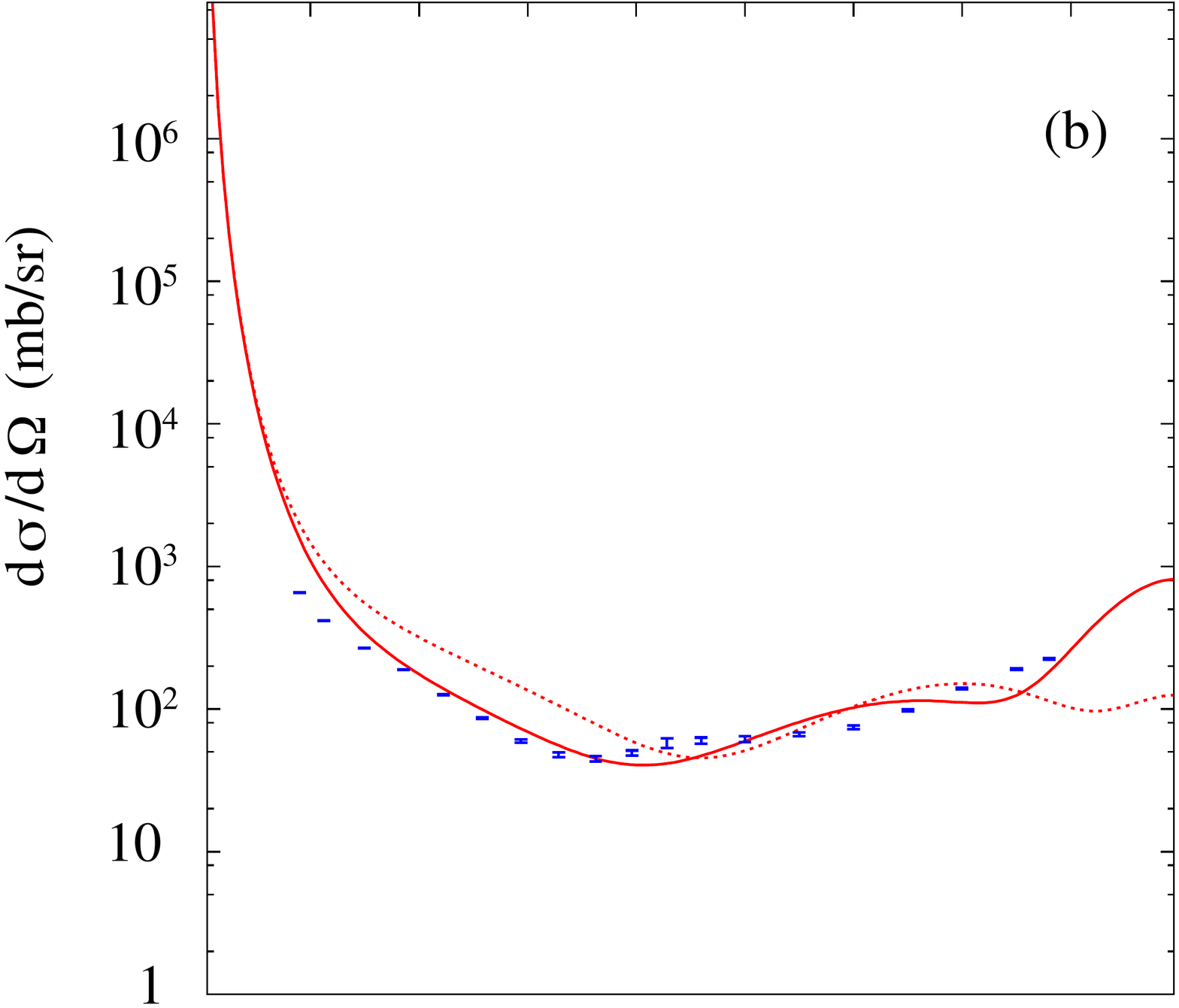} \\
\vskip -2truecm
\includegraphics[width=8.65cm]{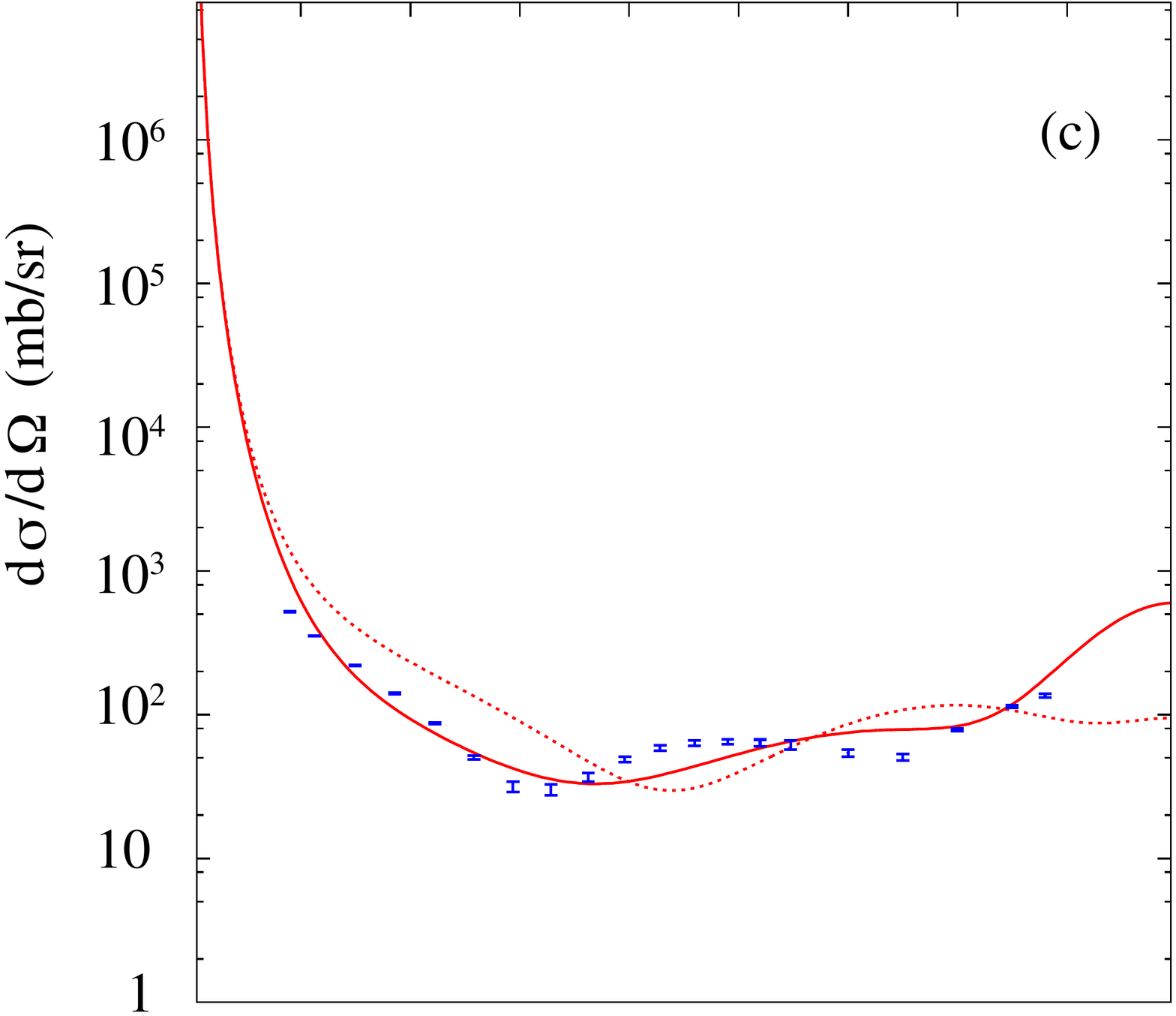} \\
\vskip -2truecm
\includegraphics[width=8.65cm]{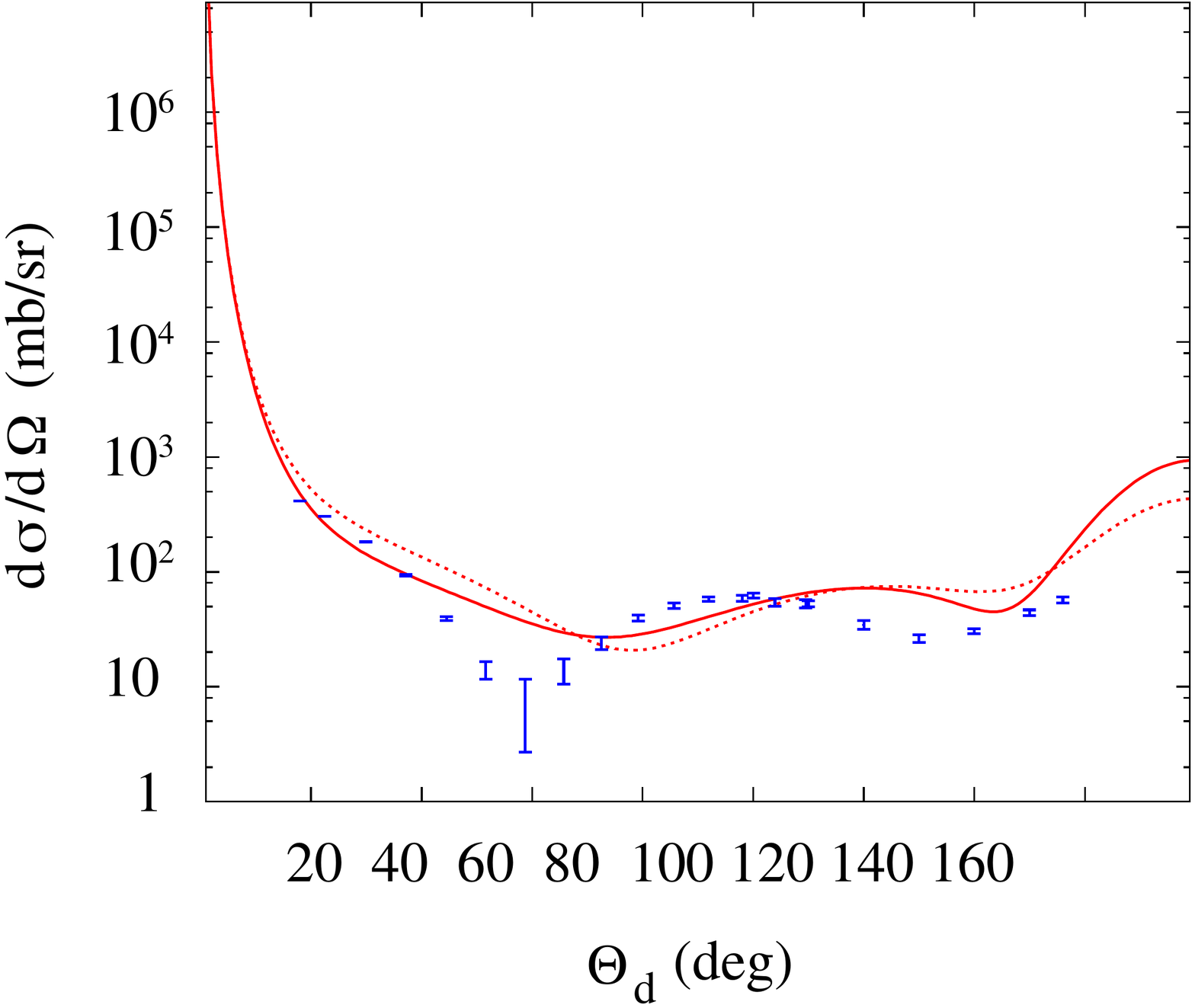} \\
\vskip -0.5truecm
\caption{(Color online) Center-of-mass frame angular distributions of the $^4$He(d,d) elastic scattering reaction calculated at four different energies (laboratory frame)
 using both the FHT(E) (dashed line) and FHT(ANC) (solid line) interactions are compared with the experimental data (symbols) \cite{differential_cross_sections_1,differential_cross_sections_2}.}
\label{Differential_cross_sections_good}
\end{figure}

The center-of-mass differential cross sections of the $^4$He(d,d) elastic scattering process have been calculated using both FHT(E) and FHT(ANC) interactions 
at four different deuteron kinetic energies in laboratory frame: 2.935 MeV, 5.961 MeV, 7.479 MeV and 11.47 MeV. 
Calculated cross sections are compared with the experimental cross sections in Fig. \ref{Differential_cross_sections_good}. 
The ability of the FHT(ANC) interaction to describe experimental data better than the FHT(E) interaction is striking.
Indeed, even low-energy differential cross sections in the GSM-CC(E) are relatively too high. This has to be compared with the phase shifts calculated with the FHT(E) interaction 
(see the dashed line in Fig. \ref{phase_shifts}). 
In particular, the $^3P_0$ phase shifts are too large in absolute value below 8 MeV, which correlates with the large differential cross section values obtained therein, 
while those calculated at 11.47 MeV are close to experimental data in the average, even though they do not follow their pattern.
While the $^3S_1$ phase shifts are correct close to 3 MeV, they rapidly become too large in absolute value afterwards.
The $^3D_2$ phase shifts also increase too quickly around 4 MeV, to remain too high afterwards due to the rather small width of 500 keV of the $2^+_1$ state.
Both these effects seem to induce wrong positions of minima and maxima in differential cross sections after 4 MeV.

Alternatively, differential cross sections calculated with the FHT(ANC) fit are always close to experimental data. 
The only discrepancy therein is that minima and maxima are slightly shifted to larger angles, and not deep enough for the first minimum at 11.47 MeV.
This cannot originate from the $^3S_1$ phase shifts, as they reproduce experimental data in the whole range of energies (see the solid line in 
Fig. \ref{phase_shifts}).
The $^3D_2$ phase shifts are also very close to experimental data, as the width of the $2^+_1$ state is larger therein, of 840 keV instead of 1.3 MeV.
Hence, discrepancies with experimental data should arise due to the $^3P_0$ and $^3D_1$ phase shifts, which are further away from the experimental data.

One can compare the differential cross sections obtained in the GSM-CC(ANC) approach with the results of the NCSMC \cite{PRL_Hupin} using realistic interactions, as they are satisfactorily reproduced in both approaches. Indeed, even though minima and maxima are at their experimental position therein, the absolute value of differential cross sections in NCSMC \cite{PRL_Hupin} and in GSM-CC(ANC) are comparable, as they are virtually exact at low energy and slightly but continually worsen when one goes to higher energy, when deuteron break-up starts to become important.
In fact, deuteron break-up is included similarly in both approaches through the use of pseudo-states in the NCSMC approach and of the intrinsic deuteron scattering states in the GSM-CC.
Moreover, the excited states of the $^4$He target are included neither in NCSMC nor in GSM-CC, and they probably have a non-negligible effect on cross sections at moderate and high energies.
As a consequence, it is possible that these features lead to the same effect in both NCSMC and GSM-CC approaches, and that they become more and more important as projectile energy increases.
\begin{center}
\begin{figure}[htb]
\includegraphics[width=9cm]{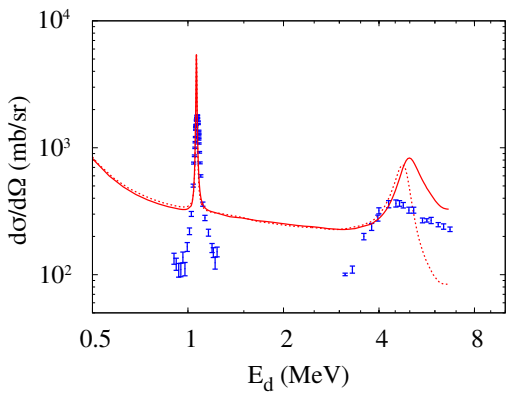} 
\vskip -0.5truecm
\caption{(Color online) Center-of-mass frame angular differential distributions for $^4$He(d,d) elastic scattering reaction at deuteron backscattered angles 164.5  and 167 degrees, 
  calculated with the FHT(E) (dotted line) and the FHT(ANC) (solid line) interactions at different deuteron kinetic energies $E_d$ in the laboratory frame. The experimental data is depicted by
symbols \cite{exp_excitation_functions_1,exp_excitation_functions_2}.}
\label{scattering_excitation_function_164p5_degrees}
\end{figure}
\end{center}
The excitation function for the $^4$He(d,d) elastic scattering reaction calculated in GSM-CC at 164.5 degrees is shown in Fig. \ref{scattering_excitation_function_164p5_degrees}.
Both FHT(E) and FHT(ANC) interactions fail to reproduce the magnitude of experimental cross section at this large angle.   
Nevertheless, the effects of the $3^+_1$ and $2^+_1$ states are clearly  visible and their widths qualitatively corresponds to that of the experiment. However, this was to be expected as the $T=0$ resonant spectrum was fitted to the experimental data. We can also see that the excitation function close to the $2^+_1$ state is better described in the GSM-CC(ANC) than in the GSM-CC(E).

The dependence of the excitation function on the first excited $1^+_2$ state of $^6$Li is not visible in calculations, whereas it is seen experimentally \cite{exp_excitation_functions_2}.
This is due to the underbinding of the calculated $1^+_2$ state which depending on the chosen parametrization of the FHT interaction lies about 1.3 to 2.6 MeV above its experimental value.
Consequently, besides having an overall too large factor of two to three in the excitation function, 
the absence of a low-lying $1^+_2$ resonance also prevents from a better reproduction of experimental data, as in the case of NCSMC calculation with realistic interactions \cite{PRL_Hupin}.

\section{Conclusion}
\label{sec5}

GSM is a model dedicated to the study of drip-line nuclei, as it incorporates both continuum and nucleon inter-correlations degrees of freedom in a unique framework.
It has been firstly devised for structure \cite{JPG_GSM_review}, and then has been successfully extended to the study of reaction cross sections with one-nucleon projectiles
within the GSM-CC approach \cite{PRC_GSM_CC_Yannen,PRC_GSM_CC_Kevin}. The inclusion of many-nucleon clusters has been far more difficult to devise, due to the internal structure of these clusters, which demands a precise treatment of their relative and center-of-mass degrees of freedom.
This has been accomplished in this paper, in the particular case of deuteron cluster, and applied to the $^4$He(d,d) elastic scattering reaction.
Even though GSM can be used in a no-core picture for the description of very light nuclei \cite{PRC_George,PRL_Fossez}, it has been chosen to use a core + valence nucleon picture
in order to be able to use GSM-CC in nuclei bearing more than 20 nucleons, which is the current limit of NCSM/RGM and NCSMC approaches.
Consequently, an effective interaction has been used therein, which recaptures the low-energy characteristics of the composite $^6$Li nucleus.

While not as microscopic as NCSMC, in which the experimental reproduction of $^4$He(d,d) observables is optimal, GSM-CC managed to quantitatively describe asymptotic normalization coefficients of $^6$Li and cross sections of the $^4$He(d,d) reaction, at the price, however, of an overbound $^6$Li ground state which points to the shortcoming of our FHT effective interaction and/or the absence of excited states and non-resonant continuum in $^4$He. The latter limitation we share with the NCSMC description of $^6$Li and $^4$He(d,d) elastic scattering. In particular, it is impossible in our case to reproduce both the correct binding energy of $^6$Li and the $^3S_1$ asymptotic normalization coefficient. It remains an open question whether this contradiction is due to the core + valence nucleon approximation or the absence of three-body forces in our approach. 

When the asymptotic normalization coefficients of $^6$Li are fitted, NCSMC and GSM-CC show comparable results for phase shifts and differential cross sections.
Hence, GSM-CC can now be used with many-nucleon clusters and we will study in a near future reactions involving many-nucleon projectiles whose composites cannot be reached in the NCSM-RGM approach.\\

We wish to thank P.~Descouvemont and G.~Hupin for useful discussions.

\end{document}